\documentclass[nojss]{jss}

\usepackage{thumbpdf,lmodern}

\usepackage{framed}

\newcommand{\TT}{\mathbf{T}}

\usepackage{amsmath,amsfonts,amssymb}

\author{Mari Myllym{\"a}ki\\Natural Resources Institute Finland (Luke)
   \And Tom\'a\v s Mrkvi\v cka\\University of South Bohemia} 
\Plainauthor{Mari Myllym{\"a}ki, Tom\'a\v s Mrkvi\v cka}

\title{Comparison of non-parametric global envelopes}
\Plaintitle{Comparison of non-parametric global envelopes}
\Shorttitle{Comparison of non-parametric global envelopes}

\Abstract{
This study presents a simulation study to compare different non-parametric global envelopes that are refinements of the rank envelope proposed by Myllym{\"a}ki et al.\ (2017, Global envelope tests for spatial processes, J. R. Statist. Soc. B 79, 381--404, doi: 10.1111/rssb.12172). 
The global envelopes are constructed for a set of functions or vectors. 
For a large number of vectors, all the refinements lead to the same outcome as the global rank envelope.
For smaller numbers of vectors the refinement playes a role, where different refinements are sensitive to different types of extremeness of a vector among the set of vectors.
The performance of the different alternatives are compared in a simulation study with respect to the numbers of available vectors, the dimensionality of the vectors, the amount of dependence between the vector elements and the expected type of extremeness. 
}

\Plainkeywords{central region, goodness-of-fit, Monte Carlo test, multiple testing, permutation test, rank envelope}

\Address{
  Mari Myllym{\"a}ki\\
  Natural Resources Institute Finland (Luke)\\
  Latokartanonkaari 9\\
  FI-00790 Helsinki, Finland\\
  E-mail: \email{mari.myllymaki@luke.fi}\\
  URL: \url{https://www.luke.fi/en/personnel/mari-myllymaki/} \\
  \emph{and}\\
  Tom\'a\v s Mrkvi\v cka \\
  Dpt. of Applied Mathematics and Informatics \\
  Faculty of Economics \\
  University of South Bohemia, \\
  Studentsk{\'a} 13 \\
  37005 \v{C}esk\'e Bud\v{e}jovice, Czech Republic \\
  E-mail: \email{mrkvicka.toma@gmail.com} \\
  URL: \url{http://home.ef.jcu.cz/~mrkvicka/} 
}

\begin{document}

\section{Introduction}

Global envelopes are useful in a number of statistical problems, e.g., for constructing central regions of sets of vectors or functions, for graphical Monte Carlo and permutation tests and for global confidence and prediction bands \citep[see e.g.][]{MyllymakiMrkvicka2020, MyllymakiEtal2017, NarisettyNair2016, MrkvickaEtal2020, MrkvickaEtal2019a, MrkvickaEtal2019b, DaiEtal2020}.
The global envelope is constructed for a set of functions or $d$-dimensional vectors, $\TT_i$, $i=1,\dots,s$.
Different global envelopes have been proposed \citep{MyllymakiEtal2017, NarisettyNair2016, MrkvickaEtal2020, MrkvickaEtal2019b}; all of them are based on a measure $M$ that orders the functions from the most extreme to the least extreme one. 
Let $M_1, \dots, M_s$ be the values of the measures for $\TT_1, \dots, \TT_s$.
For the measure, a critical value $m_{(\alpha)}$ can be determined as the largest of $M_i$ such that the number of those $i$ for which $M_i<m_{(\alpha)}$ is less or equal to $\alpha s$. The level $\alpha\in(0,1)$ is set by the user.
If the vector $\TT_i$ is among the $\alpha s$ most extreme vectors, i.e., $M_i < m_{(\alpha)}$, then it is regarded as extreme by the given measure $M_i$ at level $\alpha$.

Table \ref{tab:GEoverview} lists five different measures that are compared in this study. 
These measures are implemented in the R library GET and we refer to their detailed descriptions in \citet{MyllymakiMrkvicka2020} and in the references given in Table \ref{tab:GEoverview}.
The extreme rank length (erl), continuous rank (cont) and area rank (area) measures are all refinements of the extreme rank measure (rank), which is simply defined as the minimum of pointwise ranks of $\TT_i$ among each other. Because many $\TT_i$ can reach the same minimal rank, the given ordering of $\TT_i$ is only weak. The erl, cont and area measures break the ties in the extreme ranks and practically lead to strict ordering \citep[see][for more details]{MyllymakiEtal2017, MrkvickaEtal2019b}. 
\citet{MyllymakiEtal2017} proposed also a few global envelopes whose critical bounds depended on the estimated variances or quantiles of $\TT_i$. For comparison, we included into the simulation study the directional quantile measure (qdir), which can be recommended instead of the other proposed ones based on previous studies \citep{MyllymakiEtal2015, MyllymakiEtal2017}.

\begin{table}[ht!]
\centering
\begin{tabular}{llp{9.0cm}}
\hline
Measure & Abbr.  & Introduced in   \\ \hline
Extreme rank & rank         & \citet{MyllymakiEtal2017}  \\
Extreme rank length & erl          & \citet{MyllymakiEtal2017, NarisettyNair2016, MrkvickaEtal2020}   \\
Continuous rank & cont         & \citet{Hahn2015, MrkvickaEtal2019b} \\
Area rank & area         & \citet{MrkvickaEtal2019b}    \\
Directional quantile MAD & qdir& \citet{MyllymakiEtal2017,MyllymakiEtal2015} \\\hline
\end{tabular}
\caption{\label{tab:GEoverview} Different measures with global envelopes, their abbreviations and references. MAD stands for maximum absolute difference. 
}
\end{table}

For all the measures of Table \ref{tab:GEoverview}, a global envelope can be constructed such that it has the intrinsic graphical interpretation (IGI) \citep{MyllymakiMrkvicka2020}: the vector $\TT_i$ is outside the global envelope if and only if $M_i < m_{(\alpha)}$, and $\TT_i$ is within the global envelope if and only if $M_i \geq m_{(\alpha)}$. This holds for all $i=1, \dots, s$. 
While the IGI property is of great practical importance, in this study the focus is on the performance of the measures and visualization of the test results is not considered.

The aim of this study is to give guidance for choosing one of the measures of Table~\ref{tab:GEoverview}.
When one can afford a large number of simulations in Monte Carlo or permutation tests,
the choice of the measure plays only a minor role, because rank, erl, cont and area measures lead to an equivalent outcome for a large number of simulations. 
However, how fast the convergence happens, i.e.\ how fast ties in extreme ranks are broken along increasing $s$, depends on the situation.
On the other hand, e.g., in construction of central regions for functional data or in Monte Carlo and permutation tests where the simulations or permutations are very time consuming, the amount of available functions or vectors can be small.
Section \ref{app:simstudy} presents a simulation study on the behaviour of the measures for different numbers of simulations under different scenarios, giving guidance for the choice of the number of simulations and measures.

\section{Simulation study}\label{app:simstudy}

The choice of the measure with IGI (see Table~\ref{tab:GEoverview}) depends on a number of aspects: 1) the number of vectors which are available, 2) the dimensionality of the vectors, 3) the amount of the dependence between the vector elements, and 4) the type of extremeness which is expected.
In order to give guidance for choosing the measure, a simulation study was performed, where all of these features 1)-4) were controlled.
\begin{enumerate}
\item The considered numbers of simulated vectors $s$ were $20$, $40$, $80$, $160$, $320$, $640$, $1280$, $2560$, $5120$, $10240$, in order to cover a few functions in the central region computation up to high numbers of simulations in Monte Carlo testing.
\item The considered dimensionalities of the vectors, $d$, were $20$, $100$, $500$, $2500$, in order to cover a low-dimensional vector up to a high sampling resolution in a one-dimensional functional application or an intermediate sampling resolution in a two-dimensional functional application.
\item The simulated vectors were discretized Gaussian random processes on [0,1] with a mean of zero, a standard deviation of one and an exponential correlation function with three different values for the scale parameter, $\{0, 0.1, 1 \}$, in order to cover independence (scale 0), intermediate dependence (scale 0.1) and high autocorrelation (scale 1).
\item The first simulation in the set of simulated vectors was changed in two different ways: the function a) $5x(1-x), x\in [0,1]$ or b) $100x(1-10x)$, $x\in[0,0.1]$ was added to the Gaussian process. Case a) represents an outlier which is outlying over the whole domain $[0,1]$; this is called an {\it integral} outlier. Case b) represents an outlier which is outlying only on a small part of the domain; this is called a {\it maximum} outlier. 
\end{enumerate}

\begin{figure}[t!]
\centering
	\includegraphics[width=\textwidth]{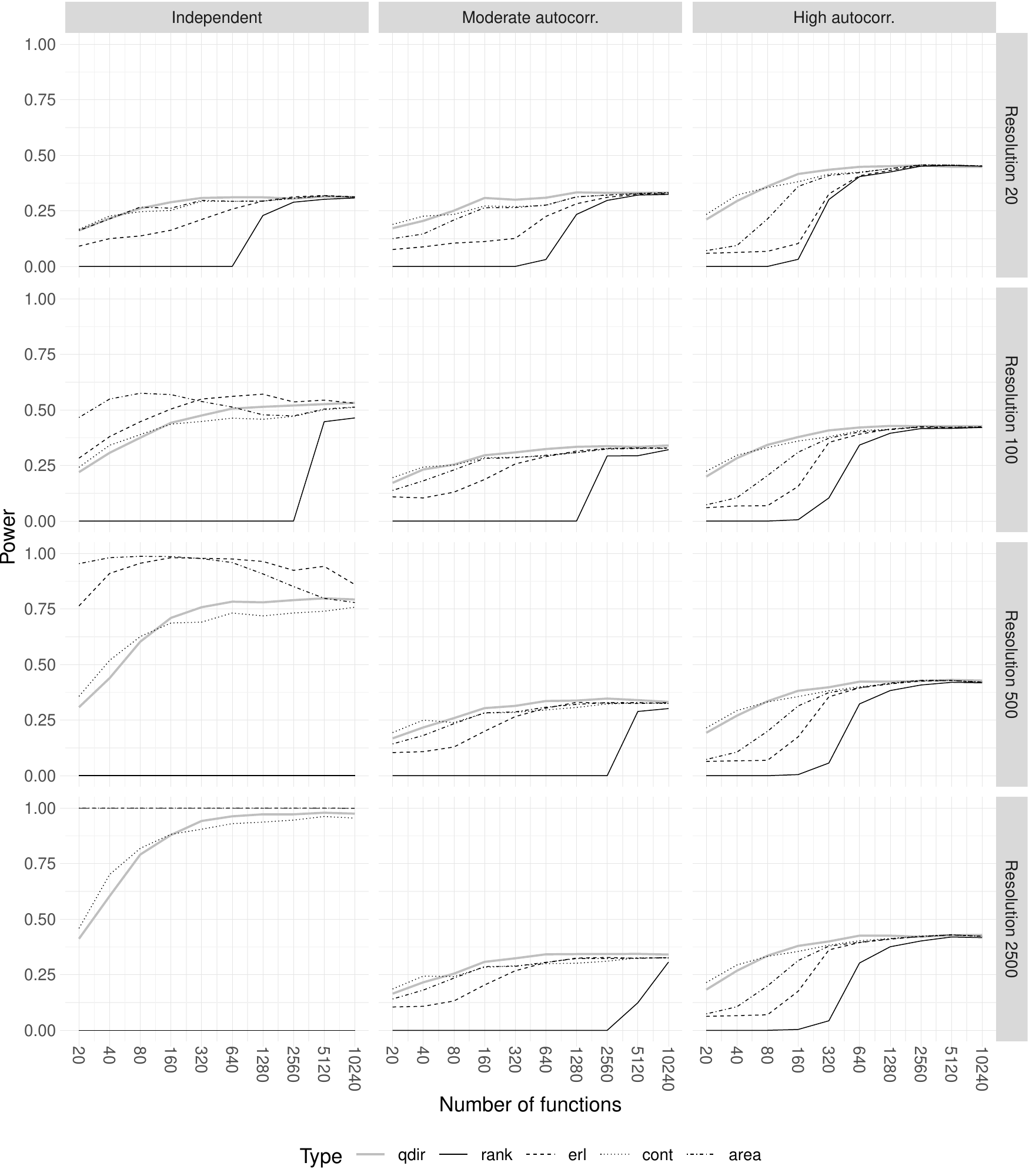}
	\caption{The estimated powers of the five different types of global envelopes (see Table~\ref{tab:GEoverview} for description) in the case of a maximum outlier for different autocorrelation in the functions (columns) and different resolutions at which the functions have been recorded (rows).}
	\label{fig:sim_int}
\end{figure}

\begin{figure}[t!]
\centering
	\includegraphics[width=\textwidth]{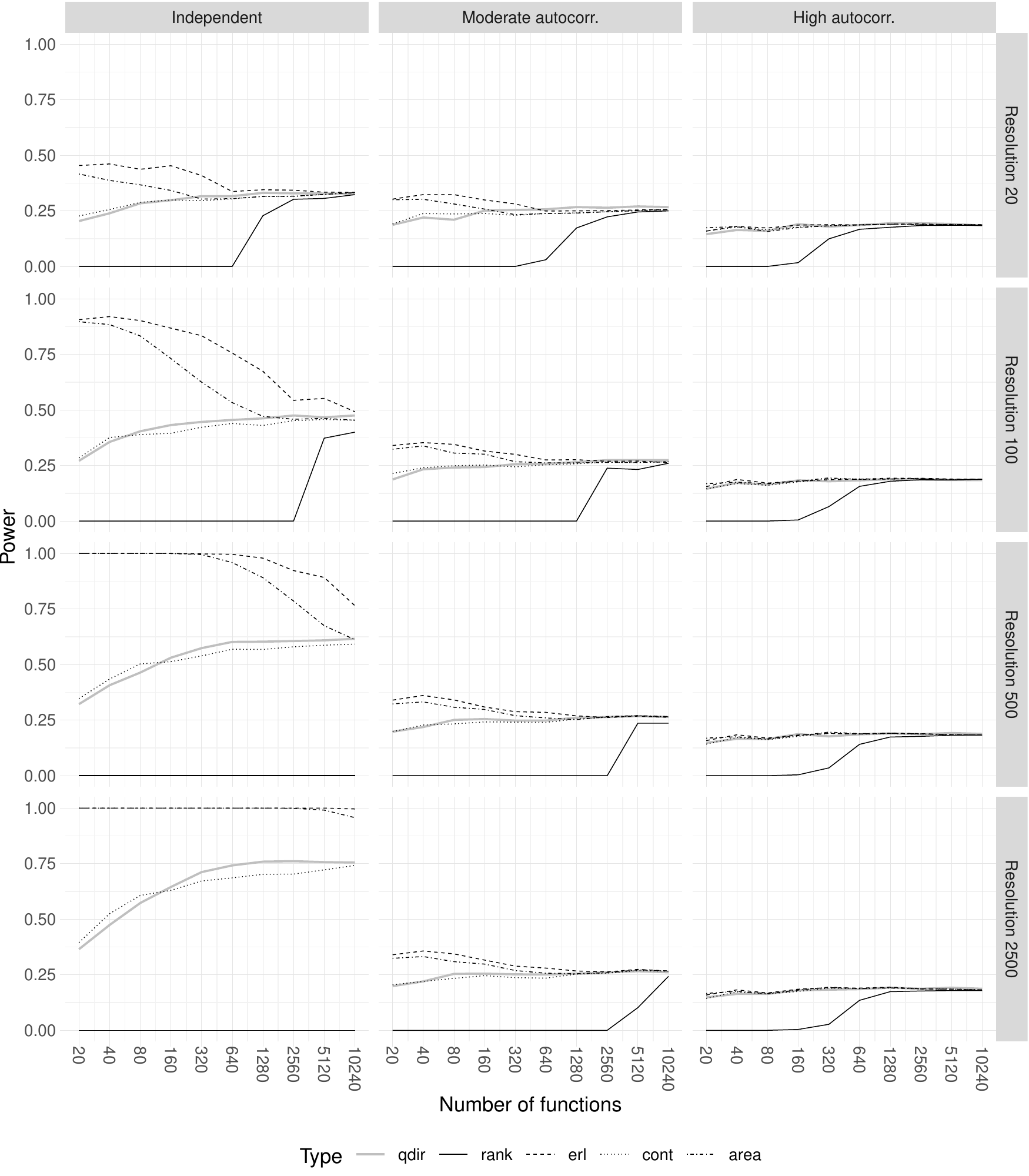}
	\caption{The estimated powers of the five different types of global envelopes (see Table~\ref{tab:GEoverview} for description) in the case of an integral outlier for different autocorrelation in the functions (columns) and different resolutions at which the functions have been recorded (rows).}
	\label{fig:sim_max}
\end{figure}

Finally, the first function was checked to see whether it was among the $\alpha s$ most extreme functions by the measures of Table~\ref{tab:GEoverview}. 
In total, 10,240 realizations of the three different Gaussian processes were generated 1000 times at the highest resolution $d=2500$, and the specifications 1)-4) were extracted from this set of functions. 
The relative numbers of positive detections (estimated powers) for each case, 1)-3), are shown in Figures~\ref{fig:sim_int} and \ref{fig:sim_max} for the integral and maximum outlier cases, respectively. These figures can be used to check which measure is the most powerful in a specific setting: for example, if one has 40 moderately auto-correlated functions with a resolution close to 100, and the expected outlier is of integral type, one can check the most interesting (powerful) measures (middle column, second row of Figure~\ref{fig:sim_int}). Or, one can check for which number of simulations or permutations would give the same result in a specific setup, e.g., for a 2D functional ANOVA model with a resolution close to 2500, a high autocorrelation, and the expected outlier is of a maximal type, one needs more than 5000 permutations (right column, fourth row of Figure~\ref{fig:sim_max}).

The following observations on the powers of the different measures can be made:
\begin{itemize}
    \item For the integral type of outlier, the erl measure had the highest power, closely followed by the area measure.
    \item For the maximum type of outlier, the best measure depended on the amount of autocorrelation and resolution. The area measure had relatively good power in all cases, even though it was beaten in the high autocorrelation case by the cont and qdir measures. 
    \item All the measures reach an equivalent outcome for large numbers of vectors. The more correlated the vectors were, the smaller was the number of vectors that was sufficient for the equivalent outcome.
    \item The power of the area and erl measures was greater for a low number of vectors than for a high number of vectors in the case of independent components and the integral outlier. 
\end{itemize}
The last point may seem surprising, but it is explained by the nature of the erl and area measures. For a low number of vectors there are many ties to break, and these ties are broken taking into account the amount (and value) of outlyingness (of integral type). On the other hand, for a high number of vectors there are less ties to break and thus the extreme rank (of maximum type) decides the order of the vectors.
	
The empirical significance levels were also checked in the case where the first function came from the same model as the rest (figure not shown), and all the obtained levels were between 0.035 and 0.067 (except for the conservative extreme rank measure, where the range was from 0 to 0.051).

\section{Conclusions}

If the number of functions or vectors is that large that there are almost no ties in the extreme ranks, then the method to break the ties does not really matter and any choice of the measure is fine. The required number of vectors depends particularly on the dependence structure of the vectors.
On the other hand, when the number of available vectors is not that large, the choice of the measure matters: The erl and area measures are typically good choices for the integral type of extremeness of the vector. For a maximum type of extremeness, the cont and qdir measures are typically the best choices, but also the area measure performs well. In a conclusion, if no particular type of extremeness is expected a priori, the area measure can be chosen as a good compromise, since it is sensitive both to the amount of outlyingness (similarly as erl) and to the value of outlyingness (similarly as cont and qdir).


\end{document}